\documentclass[12pt,a4paper]{article}
\usepackage{amsmath,amssymb,amsthm}
\usepackage{graphicx}
\usepackage{cite}

\renewcommand{\baselinestretch}{1.5}

\begin{document}

\title{{\Large \textbf{Plasma Excitations in Graphene: Their Spectral Intensity and Temperature Dependence
in Magnetic Field}}\\
}

\author{Jhao-Ying Wu$^{1}$, Szu-Chao Chen$^{1}$, Oleksiy Roslyak$^{2}$,\\ Godfrey Gumbs$^{2,\dag}$, and Ming-Fa Lin$^{1,*}$ \\
\small $^{1}$Department of Physics, National Cheng Kung University,
Tainan, Taiwan 701\\
\small  $^{2}$Department of Physics and Astronomy, Hunter College at the City University of New York,\\ \small 695 Park Avenue, New York, New York 10065, USA}

\renewcommand{\baselinestretch}{1}

\maketitle
\renewcommand{\baselinestretch}{1.5}

\begin{abstract}
In this paper, we calculated the dielectric function, the  loss function, the
magnetoplasmon dispersion relation and the temperature-induced transitions
for graphene in a uniform perpendicular magnetic field $B$. The calculations were
performed using the Peierls tight-binding model to obtain the energy band structure and
the random-phase approximation to determine the collective plasma excitation spectrum. The single-particle and collective excitations
 have been precisely identified based on the resonant peaks  in
the loss function. The critical wave vector at which
plasmon damping takes place  is clearly established.
This critical wave vector depends on the magnetic field strength
as well as the levels between which the transition takes place. The temperature effects
were also investigated. At finite temperature, there
are plasma resonances induced by the Fermi distribution function. Whether such plasmons exist is mainly determined by the field strength, temperature, and momentum. The inelastic light scattering spectroscopies could be used to verify the magnetic field and temperature induced plasmons. \vskip0.5 truecm

 \vskip0.2 truecm

\par\noindent ~~~~$^\dag$E-mail: ggumbs@hunter.cuny.edu

\par\noindent ~~~~$^\ast$E-mail: mflin@mail.ncku.edu.tw

\end{abstract}

\renewcommand{\baselinestretch}{1.4}

\newpage \renewcommand{\baselinestretch}{1.5}

 Graphene, a flat monolayer of carbon atoms with a honeycomb lattice, is the basic building block for other graphitic materials. It is famous for the linear energy dispersion around the zero Fermi energy, where the electrons can travel thousands of interatomic distances without scattering. Based on the high electron mobility, graphene is a popular candidate for the production of future nanoelectronic elements, such as ballistic transistors,$^{1-3}$ integrated circuit components,$^{4,5}$ and transparent conducting electrodes.$^{6,7}$ Other novel physics like an anomalous quantum Hall effect,$^{8,9}$ curious optical properties such as high opacity,$^{10,11}$ saturable absorption,$^{12}$ peculiar magneto-optical selection rules,$^{13}$ high thermal conductivity,$^{14,15}$ mechanical properties$^{16,17}$, and variable stacking effects$^{18,19}$ make it a possibly better candidate than any other known materials.


More fundamental physics and possible applications could be obtained by probing collective Coulomb excitations. Unlike bulk graphite$^{20}$ or bilayer graphene,$^{21}$ the low-frequency plasmon is absent in undoped monolayer graphene because of the vanishing density of states at the Fermi level. However, it was found that a rise in temperature could generate free carriers and thus intraband plasmons.$^{22}$ Accordingly, monolayer graphene could be the first undoped system which exhibits the low-frequency plasmon purely due to temperature. For doped graphene, a finite density of states at the Fermi level would lead to collective plasmon interactions at long wave lengths.$^{23-25}$ The dependence of the plasmon frequency on the momentum is described as $\omega_{P}\sim\sqrt{q}$ similar to that of a two-dimensional electron gas (2DEG).

The collective plasma excitations of graphene in a perpendicular
external magnetic field are not fully understood and still need further
 investigation. A uniform magnetic field can change the electron
density of states and consequently enhance the absorption of the
low-lying plasma excitations. Previous work based on the
 effective-mass model gives the qualitative behavior for
 magnetoplasmons in graphene.$^{26,27}$ However, a detailed quantitative
analysis has not yet been presented. The difficulty is that the Dirac equation
 can only describe the electron behavior very close to the Fermi
level (within the energy range of $\pm$0.5 eV), and the energy dispersion
 gradually becomes anisotropic outside this energy interval. In contrast, by utilizing the Peierls tight-binding model$^{28}$ and band-like matrix numerical techniques,$^{13,29}$ the entire $\pi$-magnetoelectronic
structure at realistic magnetic field strengths can be solved. The number of charge carriers per area is self conserved during
our calculations. Therefore, the accuracy of our results is not
constrained by  either the energy range or the magnetic field. In the random-phase approximation (RPA), the
 complete structure  of the dielectric function was  determined.
The single-particle and collective excitations
 can be precisely identified according to the divergences in
the loss function  $\Im m(-1/\epsilon(q,\omega))$, where $q$ is the
in-plane wave vector and $\omega$ is the frequency. It should be noted that our discussion is within the condition that $q$ is much smaller than the reciprocal-lattice vector, and the local-field effects$^{30,31}$ are neglected in our calculation. The group velocities of the magnetoplasmons in the long wavelength limit are typically positive, and then decrease to negative values as the wave vector is increased. The critical momentum for
plasmon damping to occur is clearly established.
Our calculations show that this critical wave vector has
 a strong dependence on the field strength as well as the levels between which the transition takes place. The temperature effects
were also investigated and are reported in detail below. We
found that the intra-Landau level transitions could be induced by increasing temperature at sufficiently low magnetic fields.  As a matter of fact,
the peaks corresponding to these  frequencies, which are
below the lowest inter-Landau level transition, reduce the threshold
frequency. This feature arises from the unequally spaced Landau levels (LLs) in graphene.

\begin{flushleft}
\textbf{RESULTS AND DISCUSSION}
\end{flushleft}

A uniform perpendicular magnetic field in a monolayer graphene creates many dispersionless LLs at low energy. Based on the node structure of the Landau wavefunctions, the quantum number $n^{c}$ ($n^{v}$) for each conduction (valence) LL could be determined by counting the number of zeroes; this is the same as the number labelling the $n$th unoccupied (occupied) LL above (below) the Fermi energy  $E_{F}$=0. In undoped graphene, electrons will be excited from valence LLs to conduction LLs through light absorption, for
example. However, the Coulomb interaction could affect the
the single-particle excitation mode with energy
$\hbar\omega_{ex}=E_{n^{c}}(k+q)-E_{n^{v}}(k)$ and a transferred momentum $q$. Each inter-Landau level excitation channel is labeled by ($n^{v}$, $n^{c}$) and the transition order  $\triangle n=n^{v}-n^{c}$, as depicted in Fig. 1. For example, (0,1) denotes
the transition from the highest
occupied LL to the second low unoccupied LL and has the same excitation
energy as (1,0) because of the inversion symmetry between the conduction and valence LLs. In collective mode excitations, the pair number denotes the channel with the largest contribution; this channel dominates the excitation at a large or small limit of $q$. In particular, the $n^{c,v}=0$ LL across the Fermi level may be half-filled in undoped graphene. The spin-up and spin-down states would be, respectively, at the conduction and valence bands if the Zeeman effects were considered. However, here we only deal with charge density fluctuations; there is no (0,0) excitation branch since the transition must be spin-preserved.

The single-particle excitation (SPE) spectrum and the spectrum of
collective plasmon modes due to the screened Coulomb interaction
can be well described by the behavior of the imaginary part $\epsilon_{2}$
 and the real part $\epsilon_{1}$ of the dielectric function.
In the absence of a magnetic field, $\epsilon_{2}$
is divergent at $\omega_{0}=v_{F}q$ in the asymmetric form
$1/\sqrt{\omega-\omega_{0}}$ due to the linear energy dispersion.
This is shown by the red curve in Fig. 2(a). On the other hand, the function
$\epsilon_{1}$ shown as the black curve is found to diverge as
$1/\sqrt{\omega_{0}-\omega}$. When a magnetic field is applied,
the dielectric function shows the features
displayed in Figs. 2(b) through 2(d). Each Landau level transition
 channel produces a symmetric peak in $\epsilon_{2}$ and a pair of
asymmetric peaks along with zero points in $\epsilon_{1}$. If the zero
points at which $\epsilon_{2}$ vanishes, then corresponding
to these zero points, are the frequencies of the undamped plasmon
resonances. The peak strength, determined by the Coulomb matrix elements $\upsilon_{q}|\langle n;k+q|e^{iqy}|m;k\rangle|^{2}$ shown in Eqs. (2) and (4), strongly depends on $q$. At $q$=20 ($10^{5}/cm$), the lowest excitation
channel (1,0) makes the largest contribution to the low frequency dielectric
 function as seen in Fig.\ 2(b). Other transition channels are
relatively weak. Increasing $q$ enhances the SPEs that occur
at higher frequencies, which involve the higher LLs.
The frequencies at which the highest peaks occur are dependent on $q$. Other peaks further away from these frequencies decay quickly.

The loss function, defined as  $\Im m[-1/\epsilon(q,\omega)]$, is useful for
understanding the collective excitations and the measured excitation
spectra, such as inelastic light and electron scattering spectroscopies.
Figure 3(a) shows that at $B=0$ (green line), no prominent peak exists at low $\omega$,
whereas an external magnetic field gives rise to some noticeable peaks.
These peaks may be regarded as particle-hole like or collective
plasma excitations based on the strength of the resonance, or may be
ascertained by  their  frequency. The former modes have  frequencies close to
the single particle excitations (black dashed lines) subject to
rather strong Landau damping with a finite value for $\epsilon_{2}$,
or without a zero point in $\epsilon_{1}$. On the other hand, the latter
corresponds to a zero point in $\epsilon_{1}$ and a vanishing value in
$\epsilon_{2}$ in the gap region between two SPE energies.  The smaller the
derivative  of $\epsilon_{1}$ with respect to frequency at the plasmon frequency
$\omega_p$, the higher  will be the plasmon peak, as discussed in electron
energy loss calculations for plasma excitations in the two-dimensional
electron gas as well as nanotubes.$^{32,33}$ The peak distribution, including the number, position, and intensity, strongly
depends on $q$, which demonstrates the delocalized Landau states under Coulomb
interactions. At smaller $q$, the low lying LLs dominate the excitation spectrum,
 which shifts to higher LLs when $q$ is increased. The plasmon peaks arising from the higher LL transitions have shorter heights mainly because of the
 reduced wave function overlap and the larger Landau damping out of the denser LL
 distribution. The loss function is also modulated by the field strength, as
 shown in Figs. 3(b) and 3(c). The threshold frequency decreases and the number of
  peaks increases by lowering the field strength.

The plasmon dispersion relation is shown in Fig. 4. The frequencies
of these collective modes correspond to  peak positions of
$\Im m[-1/\epsilon(q,\omega)]$  when plotted as a function of $\omega$
for fixed $q$. The minimum excitation energy  of each transition channel
 approaches the SPE energy (red dashed lines) in both the short and long
 wavelength limits where the polarization shift may be neglected. Plasmons may be
  excited over a limited range of the wave vector $q$. A characteristic behavior
  of the magnetoplasmons is that in the long wavelength limit their group
  velocity is positive; this group velocity then decreases as the wave vector is increased, and becomes
   zero at $q_{B}$ where the length scale for density fluctuations is
comparable to the cyclotron radius.  For $q>q_{B}$, the group velocity of the
 magnetoplasmons is negative and their frequencies approach those of the SPEs.
In this range of wave vectors, the magnetoplasmons encounter a loss and will be
completely damped at a large limit of $q$. From a classical point of view, the charge
density oscillations experience an external restoring force which comes
from the magnetic field. The interaction
between carriers and the magnetic field gradually becomes more important than the
 electron-electron Coulomb interaction and dominates the excitation structure
 as $q$ is increased. It is noted that the plasmon dispersion broadens and the threshold momentum of the inter-band excitations increases with the augmentation of the transition order. This is a result of the broadening of the higher Landau wavefunctions.

The critical wave vector $q_{B}$, which represents the limit momentum within which the electron wave could propagate, gradually rises with $B$ as shown in Fig. 5(a). The higher $q_{B}$ at larger $B$ corresponds to the stronger resonances in loss function (Fig. 3). The dependence of the energy-loss-peak positions on the magnetic field is shown in Fig. 5(b). The same set of excitation frequencies simply shifts for higher magnetic fields.

Temperature could induce free electrons and holes and then the intraband e-h excitations. The distribution probability of these free carriers is described by the Fermi-Dirac function. In the absence of a magnetic field, the total carrier number per area $N$ is proportional to $T^{2}$, as depicted in the inset of Fig. 6(b). The strong dependence demonstrates the $\pi$-band characteristics: zero band gap and strong wave vector-dependence. In a uniform magnetic field, $N$ would begin at a finite value that is mainly attributed to the
$n=0$ LL. When the temperature is sufficiently high, $N$ is close to that at $B=0$ because of the reduced LL spacings at higher energy. In order to observe the apparent temperature effects, the magnetic field must be sufficiently small. Here, we choose $B$=5 T and give the illustrations of temperature-dependent dielectric functions in Fig. 6 (please note that $\Gamma$ is set smaller in order to fit the narrower LL spacings). The intra-LL and inter-LL transitions coexist at finite temperature. They are marked by sharp peaks in the loss function, as shown in Fig. 7. For intra-LL transitions, the corresponding singular structures in $\epsilon$ become more pronounced and $\epsilon_{1}(\omega)$ approaches zero more slowly with an increase of the temperature, while the opposite is true for the inter-LL transitions. Consequently, the intra-LL plasmon undergoes a blue shift and its resonance peak is increased. In contrast, the inter-LL plasmon is red shifted with its height slightly reduced.

The $T$-dependent behavior of the low-frequency plasmons at different magnetic fields is shown in Fig. 8(a). To support the intra-LL plasmons, the temperature should be higher than $T_{c}$. For a smaller $B$, a lower $T_{c}$ is needed. Moreover, in a weaker magnetic field, $\omega_{p}$ shows a stronger dependence on $T$. This results from the more low-lying occupied LLs. The $T$-dependent behavior is also dominated by wave vector, as shown in Fig. 8(b). $T_{c}$ could achieve the minimum at
a critical value of $q$ and then increases; the opposite is true for the degree of $\omega_{p}-T$ dependence. This behavior may be traced to the peculiar $\omega_{p}-q$ relation.

\begin{flushleft}
\textbf{SUMMARY AND CONCLUSIONS}
\end{flushleft}

In summary, we have employed the Peierls tight-binding Hamiltonian to calculate
the electron energy bands for graphene in a uniform perpendicular magnetic field. The eigenvalues and eigenvectors were obtained efficiently by properly rearranging the base functions. With the obtained results, we were able to calculate the longitudinal wave vector and frequency-dependent
dielectric function at an arbitrary temperature. The entire LL spectrum was included in our calculations. This ensures the correctness of the dielectric function  in the RPA and
consequently the plasmon intensity  and its frequency. We presented the magnetoplasmon excitation spectra at magnetic fields which may be achieved experimentally. The exact diagonalization method could also be applicable to multi-layer graphene or
bulk graphite over a wide range of electric and magnetic fields.

The peaks in the loss function may be classified as a collective mode
or single-particle-like excitation. The former corresponds to a zero point in
$\epsilon_{1}$ at which $\epsilon_{2}$ is very small, while the latter relates to a finite value of $\epsilon_
{2}$. The peak is a function of in-plane wave vector $q$ and the magnetic field $B$.
  The critical momentum, through which the magnetoplasmon dispersion passes and
enters into the single particle mode excitation region, is clearly identified. It depends
on the transition channel and the magnetic field. The temperature effects become important when the magnetic field is sufficiently small. Extra plasmon peaks that arise from intra-LL transitions could occur in the infrared regime. The critical temperature for inducing the low-frequency modes is determined by $B$ and $q$. The numerical results
may be validated by inelastic light-scattering which has been successfully applied to the two-dimensional electron gas system in magnetic fields.$^{34,35}$

\begin{flushleft}
\textbf{METHODS}
\end{flushleft}

Monolayer graphene consists of two sublattices of A and B atoms
with a C-C bond length of b=1.42 ${\AA }$. The hopping integral between two
nearest-neighbor atoms is $\gamma_{0}$=2.598 eV. In the
tight-binding model, it is convenient to present the crystal
Hamiltonian in terms of the Bloch functions of the two periodic carbon atom
wave functions $|A\rangle$, $|B\rangle$. Here, we only take the
$2p_{z}$ orbitals to investigate the low-energy electronic structure.
The monolayer graphene is assumed to be in an ambient uniform magnetic
 field $\mathbf{B}=B\hat{z}$. The magnetic flux, the product of
the field strength and the hexagonal area, is $\Phi=
[3\sqrt{3}b^{2}B/2]/\phi_{0}$ where the flux quantum $\phi_{0}$ = hc/e = 4.1356กั1015 T/$m^{2}$. The vector potential,
which is chosen as $\mathbf{A} = Bx\hat{y}$, leads to a new periodicity
along the armchair direction. The Hamiltonian matrix element can be obtained through multiplication of the Hamiltonian matrix element in zero field by a Peierls phase.$^{28}$ Such phase has the period $R_{B}= 1/\Phi$. This performance assumes that the magnetic field changes slowly as a function of the lattice constant.$^{36}$ Therefore, the enlarged unit cell contains 4$R_{B}$ carbon atoms and the Hamiltonian matrix is 4$R_{B}$$\times$4$R_{B}$. The nearest-neighbor hopping integral associated with the extra position-dependent Peierls phase is given by

\begin{eqnarray}
\langle B_{k}|H|A_{j}\rangle&&=\gamma_0\exp{[i{\bf k}\cdot({\bf R}_{A_{j}}-{\bf R}_{B_{k}})]}
\times\exp{\{i[{2\pi\over \phi_0}\int ^{{\bf R}_{B_{k}}}_{{\bf
R}_{A_{j}}}{\bf A}\cdot d{\bf r}]\}} \cr
&&=\gamma_0t_{j}\delta_{j,k}+\gamma_0s\delta_{j,k+1} \ ,
\end{eqnarray}
where $t_{j}=\exp\{i[-k_x{b\over 2}-k_y{\sqrt{3}b\over
2}+\pi{\Phi}(j-1+{1\over 6})]\}+\exp\{i[-k_x{b\over 2}+k_y{\sqrt{3}b\over
2}-\pi{\Phi}(j-1+{1\over 6})]\}$, $s=\exp[i(-k_{x}b)]$, and ${\bf k}=(k_{x},k_{y})$ is the Bloch electron wave vector. The complete Hamiltonian matrix expression can be seen in Ref. 37. By diagonalizing the Hamiltonian, the eigenenergy $E^{c,v}$ and the wave function $\Psi^{c,v}$ are derived ($c$ and $v$ refer to the conduction and valence bands,
respectively). In the regular matrix, the eigenstates can only
 be obtained for about an $800\times 800$ matrix (for $B=400$ T).
 In order to satisfy  realistic conditions under which
 experiments are conducted ($B<50$ T for a
$6400\times 6400$ matrix), we build a band-like matrix by
rearranging the base Bloch functions.$^{13,29,37}$ This enables us to
diagonalize the large matrix and find a solution for the Landau wave functions.

Electronic excitations are characterized by the transferred
momentum $q$ and the excitation frequency
$\omega$, which determine the dielectric function. The dynamical dielectric function, calculated using the self-consistent-field
approach, is

\begin{eqnarray}
\epsilon(q,\omega)=\varepsilon_{0}-v_{q}\chi(q,\omega)
\end{eqnarray}
where $v_{q}=2\pi e^{2}/q$ is the in-plane Fourier transformation of the bare Coulomb potential energy and $\epsilon_{0}$=2.4 is the background dielectric constant for graphene. The linear response function is given by$^{38,39}$

\begin{eqnarray}
\chi(q,\omega)&=&2
\sum_{h,h'=c,v}\int_{1stBZ}\frac{dk_{x}dk_{y}}{(2\pi)^{2}}|\langle
h';\textbf{k}+\textbf{q}|e^{i\textbf{qr}}|h;\textbf{k}\rangle|^{2}
\nonumber\\
&\times &
\frac{f(E^{h'}(\textbf{k}+\textbf{q}))-f(E^{h}(\textbf{k}))}{E^{h'}(\textbf{k}+\textbf{q})
-E^{h}(\textbf{k})-(\omega +i\Gamma)}\ .
\label{polarization}
\end{eqnarray}
The Fermi-Dirac distribution is
$f(E)=1/[1+exp(E-\mu/k_{B}T)]$, where $k_{B}$ is the Boltzmann
constant and $\Gamma$ is the energy broadening which results from various deexcitation mechanisms. Furthermore, $\mu$ is the
chemical potential which is maintained at zero for any $T$ due
to the symmetry of the energy band structure about the Fermi level. The factor of two accounts for the spin degeneracy. Our discussion is constrained at the condition that $q$ is largely smaller than the reciprocal lattice vector, and thus local-field effects can be ignored.$^{31}$ In the presence
of the magnetic field, the system becomes fully quantized. The
 summation in Eq. (3) becomes a sum over all
 possible single particle transitions
 between Landau states $|m\rangle$ and $|n\rangle$. The response
function is now rewritten as:

\begin{eqnarray}
\chi(q,\omega)&=&\frac{1}{3bR_{B}\pi}
\sum_{n,m;k}|\langle n;k+q|e^{iqy}|m;k\rangle|_{q=q_{y};k=k_{y}}^{2}
\nonumber\\
&\times&
\frac{f(E_{n})-f(E_{m})}{E_{n}
-E_{m}-(\omega +i\Gamma)}\ .
\end{eqnarray}
Only $q_{y}$ and $k_{y}$ components are considered here. The calculation along the other direction of $q_{x}$ and $k_{x}$ obtains the same results. This isotropic characteristic of Landau system is the same with that in Quinn's paper.$^{40}$ Since all the $\pi$-electronic states are included in the calculations, the strength and frequency of the resonances in $\Im m(-1/\epsilon)$ can be correctly defined. Moreover, calculations on  temperature effects and highly doped systems (currently in work) are allowed.

\vskip 0.6 truecm

\noindent \textit{Acknowledgments.} This work was supported by contract \# FA 9453-07-C-0207 of AFRL, and the NSC of Taiwan under the grant No. 98-2112-M-006-013-MY4 (NCTS (south)).

\newpage
\leftline {\Large \textbf {References}}%

\begin{itemize}

\item[$^{1}$] Biel, B.; Blase, X.; Triozon. F; Roche, S. Anomalous Doping Effects on Charge Transport in Graphene Nanoribbons. \emph{Phys. Rev. Lett.} \textbf{2009}, 102, 096803-096806.

\item[$^{2}$] Wang, X.; Ouyang, Y.; Li, X.; Wang, H.; Guo, J.; Dai, H. Room-Temperature All-Semiconducting Sub-10-nm Graphene Nanoribbon Field-Effect Transistors. \emph{Phys. Rev. Lett.} \textbf{2008}, 100, 206803-206806.

\item[$^{3}$] Munoz-Rojas, F.; Fernandez-Rossier, J.; Brey, L.; Palacios, J. J. Performance Limits of Graphene-Ribbon Field-Effect Transistors. \emph{Phys. Rev. B} \textbf{2008}, 77, 045301-045305.


\item[$^{4}$] Sordan, R.; Traversi, F.; Russo, V. Logic Gates with a Single Graphene Transistor
 . \emph{Appl. Phys. Lett.} \textbf{2009}, 94, 073305-073307.

\item[$^{5}$] Yang, N.; Zhai, J.; Wang, D.; Chen, Y.; Jiang, L. Two-Dimensional Graphene Bridges Enhanced Photoinduced Charge Transport in Dye-Sensitized Solar Cells. \emph{ACS NANO} \textbf{2010}, 4, 887-894.


\item[$^{6}$] Wu, J.; Agrawal, M.; Becerril, Hector A.; Bao, Z.; Liu, Z.; Chen, Y.; Peumans, P. Organic Light-Emitting Diodes on Solution-Processed Graphene Transparent Electrodes. \emph{ACS NANO} \textbf{2010}, 4, 43-48.

\item[$^{7}$] Wang, X.; Zhi, L.; M\"{u}llen, K. Transparent, Conductive Graphene Electrodes for Dye-Sensitized Solar Cells. \emph{Nano Lett.} \textbf{2007}, 8, 323-327.


\item[$^{8}$] Zhang, Y.; Tan, Y. -W.; Stormer, H. L.; Kim, P. Experimental Observation of the Quantum Hall Effect and Berry's Phase in Graphene. \emph{Nature} \textbf{2005}, 438, 201-204.

\item[$^{9}$] McCann, E.; Fal'ko, V. I. Landau-Level Degeneracy and Quantum Hall Effect in a Graphite Bilayer. \emph{Phys. Rev. Lett.} \textbf{2006}, 96, 086805-086808.


\item[$^{10}$] Mecklenburg, M.; Woo, J.; Regan, B. C. Tree-Level Electron-Photon Interactions in Graphene. \emph{Phys. Rev. B} \textbf{2010}, 81, 245401-245405.

\item[$^{11}$] Nair, R. R.; Blake, P.; Grigorenko, A. N.; Novoselov, K. S.; Booth, T. J.; Stauber, T.; Peres, N. M. R.; Geim1, A. K. Fine Structure Constant Defines Visual Transparency of Graphene. \emph{Science} \textbf{2008}, 320, 1308-1308.


\item[$^{12}$] Sun, Z.; Hasan, T.; Torrisi, F.; Popa, D.; Privitera, G.; Wang, F.; Bonaccorso, F.; Basko, Denis M.; Ferrari, Andrea C. Graphene Mode-Locked Ultrafast Laser. \emph{ACS NANO} \textbf{2010}, 4, 803-810.


\item[$^{13}$] Ho, Y. H.; Chiu, Y. H.; Lin, D. H.; Chang, C. P.; Lin, M. F. Magneto-Optical Selection Rules in Bilayer Bernal Graphene. \emph{ACS NANO} \textbf{2010}, 4, 1465-1472.


\item[$^{14}$] Berciaud, S.; Ryu, S.; Brus, Louis E.; Heinz, Tony F. Probing the Intrinsic Properties of Exfoliated Graphene: Raman Spectroscopy of Free-Standing Monolayers. \emph{Nano Lett.} \textbf{2009}, 9, 346-352.

\item[$^{15}$] Cai, W.; Moore, Arden L.; Zhu,Y.; Li, X.; Chen, S.; Shi, L.; Ruoff, Rodney S. Thermal Transport in Suspended and Supported Monolayer Graphene Grown by Chemical Vapor Deposition. \emph{Nano Lett.} \textbf{2010}, 10, 1645-1651.


\item[$^{16}$] Scarpa, F.; Adhikari, S.;  Srikantha Phani A. Effective Elastic Mechanical Properties of Single Layer Graphene Sheets. Nanotech. \textbf{2009}, 20, 065709-065719.

\item[$^{17}$] Frank, O.; Tsoukleri, G.; Parthenios, J.; Papagelis, K.; Riaz, I.; Jalil, R.; Novoselov, Kostya S.; Galiotis, C. Compression Behavior of Single-Layer Graphenes. \emph{ACS NANO} \textbf{2010}, 4, 3131-3138.


\item[$^{18}$] Lu, C. L.; Chang, C. P.; Huang, Y. C.; Chen, R. B.; Lin, M. L. Influence of an Electric Field on the Optical Properties of Few-Layer Graphene with AB Stacking. \emph{Phys. Rev. B} \textbf{2006}, 73, 144427-144433.

\item[$^{19}$] Zhang, Y. B.; Tang, T. T.; Girit, C.;Hao, Z; Martin, M. C.; Zettl, A.; Crommie, M. F.; Shen, Y. R.; Wang, F. Direct Observation of a Widely Tunable Bandgap in Bilayer Graphene. \emph{Nature} \textbf{2009}, 459, 820-823.


\item[$^{20}$] Shyu, F. L.; Lin, M. F. Plasmons and Optical Properties of Semimetal Graphite. \emph{J. Phys. Soc. Jpn.} \textbf{2000}, 69, 3781-3784.


\item[$^{21}$] Ho, J. H.; Lu, C. L.; Hwang, C. C.; Chang, C. P.; Lin, M. F. Coulomb Excitations in AA- and AB-Stacked Bilayer Graphites. \emph{Phys. Rev. B} \textbf{2006}, 74, 085406-085413.

\item[$^{22}$] Lin, M. F.; Shyu, F. L. Temperature-Induced Plasmons in a Graphite Sheet. \emph{J. Phys. Soc. Jpn.} \textbf{2000}, 69, 607-610.


\item[$^{23}$] Shung, K. W. -K. Dielectric Function and Plasmon Structure of Stage-1 Intercalated Graphite. \emph{Phys. Rev. B} \textbf{1986}, 34, 979-993.

\item[$^{24}$] Chiu, C. W.; Lee, S. H.; Chen, S. C.; Lin, M. F. Electronic Excitations in Doped Monolayer Graphenes. \emph{J. Appl. Phys.} \textbf{2009}, 106, 113711-113716.

\item[$^{25}$] Wunsch, B.; Stauber, T.; Sols, F.; Guinea, F. Dynamical Polarization of Graphene at Finite Doping. \emph{New J. phys.} \textbf{2006}, 8, 318-332.


\item[$^{26}$] Rold$\acute{a}$n, R.; Fuchs, J.-N.; Goerbig, M.O. Collective Modes of Doped Graphene and a Standard Two-Dimensional Electron Gas in a Strong Magnetic Field: Linear Magnetoplasmons versus Magnetoexcitons. \emph{Phys. Rev. B} \textbf{2009}, 80, 085408-085413.

\item[$^{27}$] Berman, O. L.; Gumbs, G.; Lozovik, Y. E. Magnetoplasmons in Layered Graphene Structures. \emph{Phys. Rev. B} \textbf{2008}, 78, 085401-085405.

\item[$^{28}$] Saito, R.; Dresselhaus, G.; Dresselhaus, M. S. \emph{Physical Properties of Carbon Nanotubes}; Imperial College Press: London, 1998.

\item[$^{29}$] Lai, Y. H.; Ho, J. H.; Chang, C. P.; Lin, M. F. Manetoelectronic Properties of Bilayer Bernal Graphene. \emph{Phys. Rev. B} \textbf{2008}, 77, 085426-085435.

\item[$^{30}$] Wiser, N. Dielectric Constant with Local Field Effects Included. \emph{Phys. Rev.} \textbf{1963}, 129, 62-69.

\item[$^{31}$] Tudorovskiy, T.; Mikhailov, S. A. Intervalley Plasmons in Graphene. \emph{Phys. Rev. B} \textbf{2010}, 82, 073411-073414.

\item[$^{32}$] Horing, N. J. M.; Tso, H. C.; Gumbs, G. Fast-Particle Energy Loss in the Vicinity of a Two-Dimensional Plasma. \emph{Phys. Rev. B} \textbf{1987} 36, 1588-1594.

\item[$^{33}$] Balassis, A.; Gumbs, G. Plasmon Instability in Energy Transfer from a Current of Charged Particles to Multiwall and Cylindrical Nanotube Arrays Based on Self-Consistent Field Theory. \emph{Phys. Rev. B} \textbf{2006}, 74, 045420-045427.

\item[$^{34}$] Richards, D. Inelastic Light Scattering from Inter-Landau Level Excitations in a Two-Dimensional Electron Gas. \emph{Phys. Rev. B} \textbf{2000}, 61, 7517-7525.

\item[$^{35}$] Eriksson, M. A.; Pinczuk, A.; Dennis, B. S.; Simon, S. H.; Pfeiffer, L. N.; West, K. W. Collective Excitations in the Dilute 2D Electron System. \emph{Phys. Rev. Lett.} \textbf{1999}, 82, 2163-2166.

\item[$^{36}$] Luttinger, J. M. The Effect of a Magnetic Field on Electrons in a Periodic Potential. \emph{Phys. Rev.} \textbf{1951}, 84, 814-817.

\item[$^{37}$] Ho, J. H.; Lai, Y. H.; Chiu, Y. H.; Lin, M. F. Landau Levels in Graphene. \emph{Physica E} \textbf{2008}, 40, 1722-1725.

\item[$^{38}$] Ehrenreich, H.; Cohen, M. H. Self-Consistent Field Approach to the Many-Electron Problem. \emph{Phys. Rev.} \textbf{1959}, 115, 786-790.

\item[$^{39}$] Shung, Kenneth W. -K. Dielectric Function and Plasmon Structure of Stage-1 Intercalated Graphite. \emph{Phys. Rev. B} \textbf{1986}, 34, 979-993.

\item[$^{40}$] Chiu, K.~W.; Quinn, J.~J. Plasma Oscillations of a Two-Dimensional Electron Gas in a Strong Magnetic Field. \emph{Phys. Rev. B} \textbf{1974}, 9, 4724-4732.

\end{itemize}

\newpage \centerline {\Large \textbf {FIGURE CAPTIONS}}

\vskip0.5 truecm

Figure 1. Schematic diagram showing the inter-Landau-level transitions.

\vskip0.5 truecm

Figure 2. The real and imaginary parts of the dielectric function $\epsilon(q,\omega)$ as a function of frequency $\omega$ for selected wave numbers $q$. (a)  B=0 and $q=40$, (b) B=40 T and $q= 20$, c) B=40 T and $q= 40$, and (d) B=40 T and $q= 60$. The wave number $q$ is measured in units of $10^{5}cm^{-1}$, here and elsewhere in this paper.

\vskip0.5 truecm

Figure 3. The loss functions at $B$=40 T for different values of the in-plane wave vector $q$ in (a), and for different field strengths in (b) and (c).

\vskip0.5 truecm

Figure 4. Plot of magnetoplasmon frequency as a function of the in-plane wave vector $q$.
Only several of the lowest branches are shown. We choose a magnetic field of $B$=40 T.

\vskip0.5 truecm

Figure 5. (a) Wave number $q_{B}$, where the magnetoplasmon group velocity is zero, as a function of the magnetic field. (b) Magnetoplasmon frequency for $q=20$ as a function of the magnetic field.

\vskip0.5 truecm

Figure 6. Plots of (a) the real part and (b) the imaginary part of the dielectric function as a function of the frequency for a chosen magnetic field and a in-plane wave vector $q$. Three different temperatures were chosen in these calculations. The inset in (b) shows  the number of carriers per unit area as a function of the temperature for two chosen magnetic fields.

\vskip0.5 truecm

Figure 7. The loss function as a function of the frequency for a chosen magnetic field and a in-plane
wave vector. Results are presented for four temperatures.

\vskip0.5 truecm

Figure 8. The temperature-dependent intra-LL plasmon frequencies at (a) different magnetic fields and (b) different momentum.

\newpage

\end{document}